\documentclass[12pt]{article}
\usepackage{graphicx}
\usepackage{amssymb}

\def\bq{\begin{equation}}
\def\eq{\end{equation}}
\def\ba{\begin{array}}
\def\ea{\end{array}}
\def\e{\varepsilon}

\begin{document}
\begin{titlepage}
\begin{center}
{\huge
Experimental and theoretical study of the passively
mode-locked Ytterbium-doped double-clad fiber laser}
\end{center}
\vspace{1.5cm}

\noindent
Herv\'e Leblond$\rm ^{(a)}$, Mohamed Salhi$\rm ^{(a)}$, Ammar Hideur$\rm ^{(b)}$,
 Thierry Chartier$\rm ^{(b)}$, Marc Brunel$\rm ^{(b)}$ and Fran\c{c}ois Sanchez$\rm
 ^{(a)}$\\\vspace{1cm}

 {\it \noindent
(a) Laboratoire POMA, UMR 6136, Universit\'e d'Angers, 2 Bd
Lavoisier, 49000 Angers, France\\
 (b) Groupe d'Optique et
d'Optronique, CORIA, UMR 6614, Universit\'e de Rouen, Avenue de l'Universit\'e,
 site Universitaire du
Madrillet, B.P. 12, 76801 Saint-\'Etienne du Rouvray, France}

\vspace{2.5cm}

{\bf Abstract}\vspace{5mm}

We consider an Yb-doped double-clad fiber laser in a unidirectional ring cavity containing a polarizer placed between
two half-wave plates. Depending on the orientation of the phase plates, the laser operates in continuous, Q-switch,
mode-lock or unstable self-pulsing regime. An experimental study of the stability of the mode locking regime is
 realized versus the orientation of the half-wave plates. A model for the stability of  self-mode-locking and
  cw operation  is
 developed starting from two coupled nonlinear Schr\"odinger equations in a gain medium. The model is reduced to a
  master equation in which the coefficients are explicitly dependent on the orientation angles of the phase plates.
   Analytical solutions are given together with their stability versus the angles.

\end{titlepage}
\section{Introduction}

Self-mode-locking has been successfully obtained in both low power and double-clad fiber lasers using Additive-Pulse
 Mode locking technique through nonlinear polarization rotation \cite{1}-\cite{6}. The experimental configuration generally consists
 in a ring cavity containing a polarizing isolator placed between two polarization controllers. The basic principle is the
  following. The polarization state evolves nonlinearly during the propagation of the pulse in the fiber due to
  the combined effects of self-phase modulation and cross-phase modulation induced on the two orthogonal polarization
  components, both resulting from the optical Kerr effect. A polarization controller is adjusted at the output of the
  fiber such that the polarizing isolator lets pass the central intense part of the pulse but blocks the low-intensity
   pulse wings. This technique of nonlinear polarization rotation for passive mode locking leads to stable, self-starting
   pulse trains. Theoretical approaches to this problem can be divided into two categories. The first one has been proposed by
    Haus {\it et al} \cite{7} and consists in writing a phenomenological scalar equation (the master mode-locking equation) assuming that all
     effects per pass are small. The model includes the group velocity dispersion (GVD), the optical Kerr nonlinearity and a gain medium.
      No birefringence is taken into account. Although  the mode-locking properties of the laser can be described through this approach
      for
      positive and negative GVD,  the stability of the mode-locked solutions as a function of
       the orientation of the polarization controllers cannot be described in this way. On the other hand Kim {\it et al} \cite{8} use an approach based
        on two coupled nonlinear Schr\"odinger equations. The medium is assumed to be birefringent.
        The GVD
         together with the Kerr effect are taken into account. The orientation angles of the eigenaxis of the fiber at both sides
         of the polarizer are explicitly taken into account through a periodic perturbation. Numerical simulations show that
          mode-locking can be achieved for negative GVD. To our best knowledge, none of the models take
          into account all the characteristics of the fiber (birefringence, GVD, gain, optical Kerr effect)
           and the action of the orientation of the polarization controllers.

The aim of this paper is to investigate experimentally and theoretically the mode-locking properties of the Yb-doped double-clad fiber
 laser. Mode-locking is achieved through nonlinear polarization rotation in a unidirectional ring cavity containing a polarizer
  placed between two half-wave plates. This paper is organized as follows. Section II is devoted to the experimental results. We first
   briefly recall the operating regime of the laser as a function of the orientation of the two half-wave plates. A stability diagram
   of the mode-locked laser is presented \cite{9}. The parameters required in the modelling section are measured. The birefringence
   of the double-clad fiber is first determined using the magneto-optic method \cite{10}. The total dispersion has been
   estimated from other experiments where compression of the pulses was realized
 \cite{11,11bis}. In section \ref{III}  the
    stability of the mode-locking regime is theoretically investigated. The analysis is based on the coupling
     between the two linearly polarized eigenstates
    of the fiber \cite{8}\cite{12}. The propagation problem is modelled through two nonlinear Schrodinger equations in a gain medium where the
     GVD, the nonlinear coefficient and the gain are assumed small over the cavity length. The eigenaxis of
     the fiber at each side of the polarizer are referenced with respect to the polarizing axis. These angles can be adjusted with
      two half-wave plates and  will thus be considered as variable in the analysis. The polarizer is  treated as a periodic
      perturbation acting as a projecting operator. Using the expressions derived in the propagation problem, we calculate the
       electric field amplitude just after the polarizer as a function of the corresponding amplitude at the previous round-trip.
        The round-trip number is then considered as a continuous variable, which allows us to use a multiscale analysis approach.
         The final equation is a complex Ginzburg-Landau equation whose coefficients explicitly involve the orientation angles
          of the half-wave plates. Explicit solutions are given and their stabilities are discussed. A stability diagram of
          the mode-locked regime is calculated and compared to the experimental diagram.

\section{Experimental results}

\begin{figure}[hbt!]
\begin{center}
\includegraphics[width=9cm]{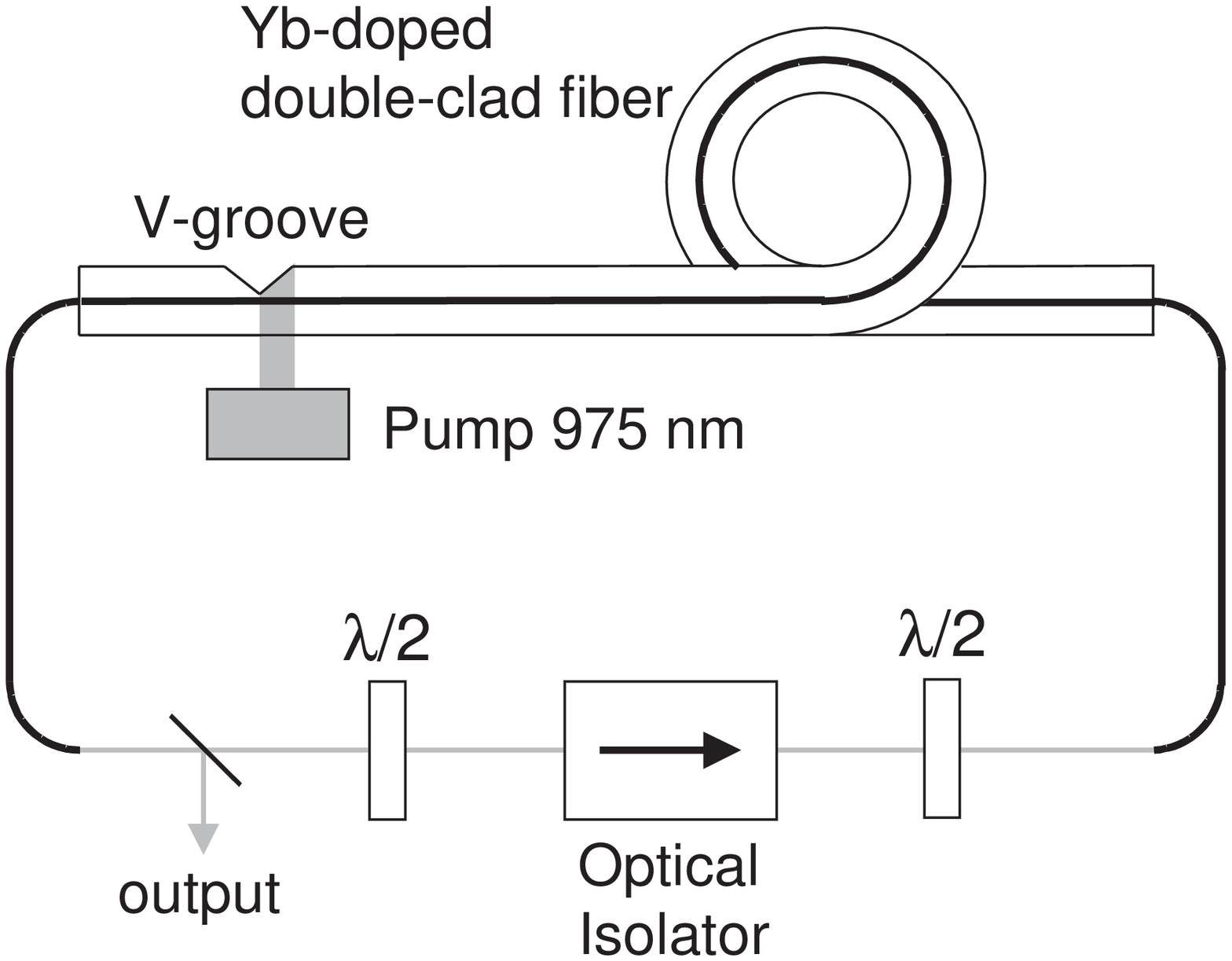}
\caption{\footnotesize   Experimental setup.}
\label{figs1}
\end{center}
\end{figure}

In this section we briefly recall the results obtained with an Yb-doped double-clad fiber laser in a unidirectional
 ring cavity operating at 1.08 $\rm \mu m$ \cite{9}. A polarizing isolator is placed between two half-wave plates. Although this
 configuration is known to generate ultrashort pulses in low power fiber lasers, it can also lead to very different
  regimes depending on the orientation of the polarization elements, such as cw operation, Q-switching, mode-locking,
   or unstable emission regime. We thus perform a systematic study of the operating regime as a function of the relative
    orientation of the two half-wave plates. The experimental setup is shown in figure \ref{figs1}.
    The 4 meter long Yb-doped double-clad
     fiber is side-pumped using the v-groove technique \cite{3}. The launching efficiency is about 70 percent. This pumping
     scheme is very convenient for ring laser applications because the two fiber ends remain free. The fiber core diameter
      is 7 $\rm \mu m$ allowing a single-mode laser signal propagation. The inner cladding is a $125\times 125\;\;\rm \mu m^2$
       square leading to a
       multimode pump propagation. The geometry of the inner cladding ensures an efficient pump absorption. Indeed,
        the pump power is almost completely absorbed along the doped fiber. In addition, multimode propagation is required
        for high injection efficiency of large area laser diodes. Two single-mode fibers at 1 $\rm \mu m$ are spliced at both ends of
         the double-clad fiber leading to a total fiber length of about 8.5 m, the cavity length is about 9 m. Fiber ends
         are angle-cleaved in order to avoid any Fresnel reflection which could generate a standing-wave operating regime.
          The splicing losses are less than 0.1 dB. The doped fiber is pumped at 975 nm with a 4 W semiconductor laser.
           A bulk polarizing optical isolator is used to obtain a travelling wave laser. The latter is placed between
          two half-wave plates at 1.08 $\rm \mu m$. Each plate can be rotated in a plane perpendicular to the propagation direction.
           A partially reflecting mirror  ensures an output coupling of about 90 percent. The output beam
            is analyzed through different means. A fast photodiode is used when the laser is cw, Q-switched or unstable.
            In the regular mode-locking regime, we use an optical autocorrelator to measure the duration of the pulses.
            Note that there is no compression system in the cavity. An optical spectrum analyzer (Advantest) together with
             a microwave spectrum analyzer (Tektronix) are also used when needed. Finally, a powermeter allows us to measure
              the average output power.

\begin{figure}[hbt!]
\begin{center}
\includegraphics[width=9cm]{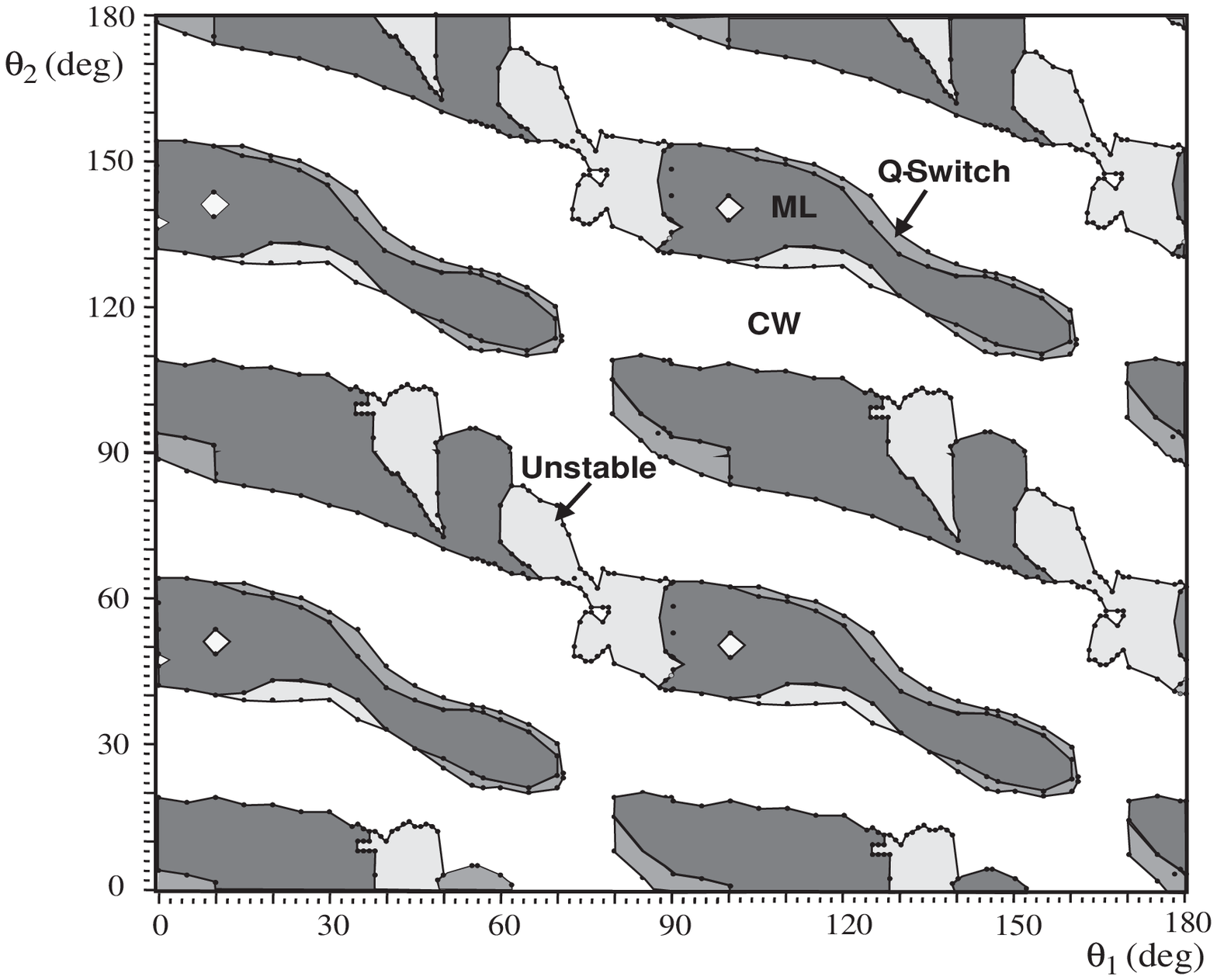}
\caption{\footnotesize  Operating regime of the laser as a function of the orientation of the two half-wave plates $\theta_1$ and $\theta_2$.
 The pump power was 2.75 W. ML stands for mode-locked regime and cw for continuous wave regime.}
\label{figs2}
\end{center}
\end{figure}

We investigate now the operating regime of the laser as a function
of the orientation of the two waveplates. $\theta_1$ ($\theta_2$)
is the orientation of one eigenaxis of the first (second) half-wave
plate referenced to the input (output) polarizer of the optical
isolator. The angle between the input and output polarizer of the
isolator is $\rm 45$ degrees. For a given pumping power we record the
evolution of the output
 intensity versus $\theta_1$ and $\theta_2$. Practically, we fix $\theta_1$ and measure the particular values of $\theta_2$
 for which a change in the operating regime occurs. It is convenient to represent these results in the plane $(\theta_1,\theta_2)$.
 It  allows to easily identify the output regime of the laser. The experimental results, obtained for a pumping power
   of about 2.75 W, are summarized in figure \ref{figs2}.
    We have experimentally observed that the different output regimes are periodic
   versus $\theta_1$ and $\theta_2$ with a period of  90 degrees.
    Figure \ref{figs2} shows that the laser essentially operates in cw, mode-lock or unstable regime while the Q-switch behavior appears near
     the boundary between cw and mode-lock regimes. Note that since the operating regime depends on the pumping rate,
      the mapping of figure \ref{figs2} is pump power dependent. Typically, when the pump power decreases (increases), the ranges of
       mode-locking decrease (increase).

\begin{figure}[hbt!]
\begin{center}
\includegraphics[width=9cm]{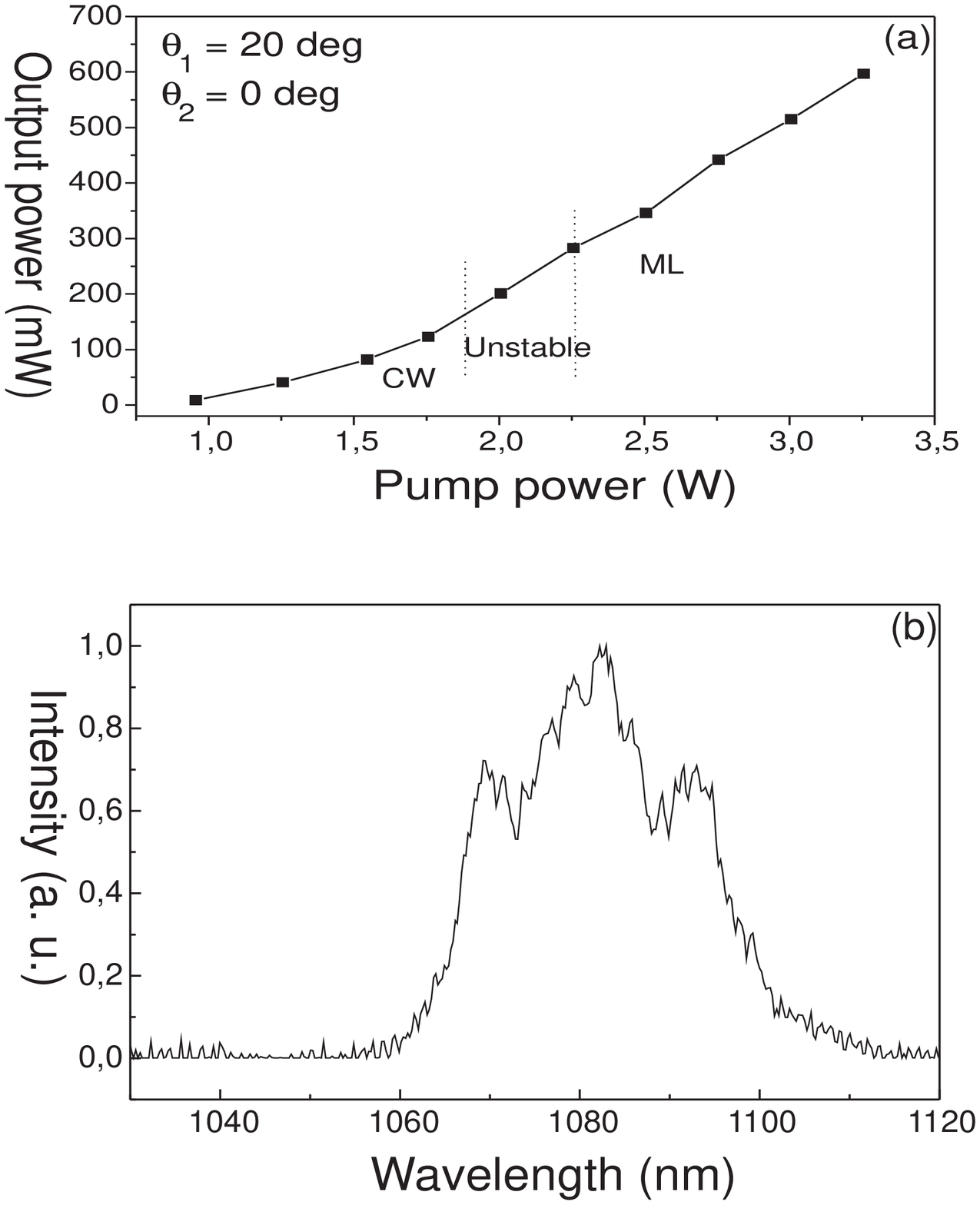}
\caption{\footnotesize   $\theta_1 = \rm 20\, deg$ and $\theta_2 = \rm 0\, deg$. (a) laser characteristic and (b) typical optical spectrum in
 the mode-locked regime.}
\label{figs3}
\end{center}
\end{figure}

The different dynamical behaviors observed in our experiment have been discussed in details in reference \cite{9}.
 Here, we concentrate our attention on
the mode-locked operating regime. In figure \ref{figs3} we give the results obtained for $\theta_1 = \rm 20\, deg$
and $\theta_2 =\rm  0\, deg$.
 In this case, the laser is first cw just above threshold (figure \ref{figs3}a). While the pump power is increased, an irregular
  self-pulsing regime appears which finally becomes a regular self-mode-locking regime. As previously discussed in the introduction,
   the basic principle responsible for the mode-locking is well known in such configurations. It is based on Kerr effect through
   nonlinear polarization rotation. This technique is also called Polarization-Additive Pulse Mode-locking. The repetition rate
    of the pulses corresponds to the free spectral range of the cavity, about 20 MHz in our case. Auto-correlation traces showed
     a coherence spike of about 150 fs on a longer pedestal of about 60 ps \cite{6}. A typical spectrum in the mode-locking range is given
      in figure \ref{figs3}b.
       The mode-locked spectrum bandwidth is more than 25 nm. This spectrum suggests that shorter pulses can be achieved
       by a suitable dispersion compensation. We have recently shown that femtosecond pulses are obtained with a grating pair inserted
        in the cavity \cite{6}. In the present  mode-locked regime, we measure 600 mW output power for 3.25 W pump power, which corresponds to high
        energy pulses of about 30 nJ.

Consider now the measurements of parameters required in the theoretical section. We first measure the beating length $L_B$
 of the double-clad fiber using the magneto-optic method \cite{10}. The method consists to apply a sinusoidal magnetic field produced
 by a coil in a small portion of the fiber and to move the coil along the fiber axis. The fiber is placed between an analyzer and
  a polarizer. This system is then illuminated and analyzed in transmission. The detected output power is a periodic signal versus
   the coil displacement. The period in the spatial domain coincides with the beating length of the fiber. We found $L_B = 50\rm\; cm$
 leading to a birefringence parameter $K = 1/( L_B) = 2\;\rm m^{-1}$. We must also  consider the birefringence of the undoped fibers.
  It has been measured to be of about $1\;\rm  m^{-1}$. In the following we will consider an average value
  of the birefringence of about $1.5\;\rm  m^{-1}$. A second important parameter is the GVD.
   It has been estimated
 from compression experiments \cite{11bis}. We find $\beta_2 = 0.026\;\rm  ps^2/m$.

\section{Theory}\label{III}
The aim of this section is to develop a theoretical model able to describe the mode-locking properties of the Yb-doped fiber laser
as a function of the orientation of the two half-wave plates. For that, we consider a birefringent fiber in a unidirectional ring cavity
with an intracavity polarizer. The orientation of the eigenaxis of the fiber at each sides of the polarizer are unspecified and will
 be taken as variable parameters. This is equivalent to considering that the eigenaxis of the fiber are fixed and two half-wave
  plates are used to adjust their orientations with respect to the polarizer.

\subsection{Derivation of a master equation}
\subsubsection{Propagation in the fiber}
The starting point are the equations giving the evolution of the two polarization components in a gain medium with Kerr nonlinearity
and GVD. In the framework of the eigenaxis of the birefringent fiber moving at the
group velocity,
the pulse envelope evolution is described by the following system \cite{8}\cite{12}:
\bq i\partial_zu-K u-\frac{\beta_2}2\partial_t^2u
+\gamma\left(u\left|u\right|^2+Au\left|v\right|^2+Bv^2u^\ast\right)=
ig\left(1+\frac1{\omega_g^2}\partial_t^2\right)u,\label{1}\eq
\bq i\partial_zv+K v-\frac{\beta_2}2\partial_t^2v
+\gamma\left(v\left|v\right|^2+Av\left|u\right|^2+Bu^2v^\ast\right)=
ig\left(1+\frac1{\omega_g^2}\partial_t^2\right)v,\label{2}\eq
where $\partial_t$ denotes the partial derivative operator $\frac{\partial}{\partial t}$.
$\beta_2$ (in $\rm ps^2/m$) is the GVD coefficient. $K$ (in $\rm m^{-1}$)
is the birefringent parameter and is related to the $x$ and $y$ refractive indexes through the
relation $K = \pi(n_x-n_y)/\lambda$, where lambda is the optical wavelength.
$\gamma = 2\pi n_2/(\lambda A_{eff})$ is the nonlinear coefficient, $n_2$ (in $\rm m^2/W)$
 is the nonlinear index coefficient and $A_{eff}$ (in $\rm m^2$) the effective core area of the fiber.
 $A$ and $B$ are the dielectric coefficients. In isotropic media, $A = 2/3$ and $B = 1/3$ \cite{12}. This is the case with our
 silicate fibers. $g$ is the gain coefficient (in $\rm m^{-1}$) and $\omega_g$
  is the spectral gain bandwidth (in $\rm ps^{-1}$).
   At first order the gain coefficient is fixed by the fact that  it compensates the losses.
Note that polarization mode dispersion is not taken into account since the cavity length is short (about 9 m).
For numerical simulations we will use the following values for the parameters: $K = 1.5\rm \; m^{-1}$,
$ \beta_2 = 0.026\;\rm ps^2/m$, $L = 9\;\rm m$, $\gamma = 3\cdot 10^{-3}\;\rm  W^{-1} m^{-1}$ and $\omega_g = 10^{13}\;\rm s^{-1}$.

We now assume that the GVD $\beta_2$, the nonlinear coefficient $\gamma$, the gain filtering
$\rho={g}/{\omega_g^2}$ are small over one round-trip of the cavity. We introduce a small parameter $\e$ and replace
these quantities by $\e\beta_2$, $\e\gamma$ and $\e\rho$. We then look for solutions of the system (\ref{1}-\ref{2}) under the
form of a power series in $\e$, as
\bq  u=u_0+\e u_1+O\left(\e^2\right),\eq
\bq  v=v_0+\e v_1+O\left(\e^2\right).\eq
At order $\e^0$ it yields
\bq u_0=\tilde u_0e^{(g-iK)z},\eq
\bq v_0=\tilde v_0e^{(g+iK)z},\eq
where $\tilde u_0$ and $\tilde v_0$ do not depend on $z$, {\it ie} depend on the time variable $t$ only.

Making the transformation
\bq u_1=\tilde u_1(z,t)e^{(g-iK)z},\eq
\bq v_1=\tilde v_1(z,t)e^{(g+iK)z},\eq
the equations obtained at order $\e$ can be integrated with regard to $z$
to yield, for the component $u$:
\bq \ba{rl}
\displaystyle\tilde u_1=&
\displaystyle\tilde u_1(0)+z\left(\rho-\frac{i\beta_2}2\right)\partial_t^2\tilde u_0\\
&\displaystyle
+i\gamma\left(\tilde u_0\left|\tilde u_0\right|^2
+A\tilde u_0\left|\tilde v_0\right|^2\right)\frac{e^{2gz}-1}{2g}
+i\gamma B\tilde v_0^2\tilde u_0^\ast\frac{e^{(2g+4iK)z}-1}{2g+4iK}.\ea\label{3}\eq
To get the complete expression for $u(z)$, we notice that
\bq u(0)=\tilde u_0+\e \tilde u_1(0)+O\left(\e^2\right).\eq
Then the two components of the wave amplitude can be written as
\bq \ba{rl}
\displaystyle u(z)=&
\displaystyle u(0)e^{(g-iK)z}+\e\biggl[ z\left(\rho-\frac{i\beta_2}2\right)\partial_t^2u(0)
\\
&\displaystyle+i\gamma\left(u(0)\left|u(0)\right|^2
+A u(0)\left| v(0)\right|^2\right)\frac{e^{2gz}-1}{2g}
\\
&\displaystyle\hspace{1.5cm}
+i\gamma B\tilde v(0)^2\tilde u(0)^\ast\frac{e^{(2g+4iK)z}-1}{2g+4iK}\biggr]e^{(g-iK)z}
+O\left(\e^2\right),\ea
\label{4}\eq
\bq \ba{rl}
\displaystyle v(z)=&
\displaystyle v(0)e^{(g+iK)z}+\e\biggl[ z\left(\rho-\frac{i\beta_2}2\right)\partial_t^2v(0)
\\
&\displaystyle+i\gamma\left(v(0)\left|v(0)\right|^2
+A v(0)\left| u(0)\right|^2\right)\frac{e^{2gz}-1}{2g}
\\
&\displaystyle\hspace{1.5cm}
+i\gamma B\tilde u(0)^2\tilde v(0)^\ast\frac{e^{(2g-4iK)z}-1}{2g-4iK}\biggr]e^{(g+iK)z}
+O\left(\e^2\right).\ea
\label{5}\eq

\subsubsection{Polarizer}
\begin{figure}[hbt!]
\begin{center}
\includegraphics[width=6cm]{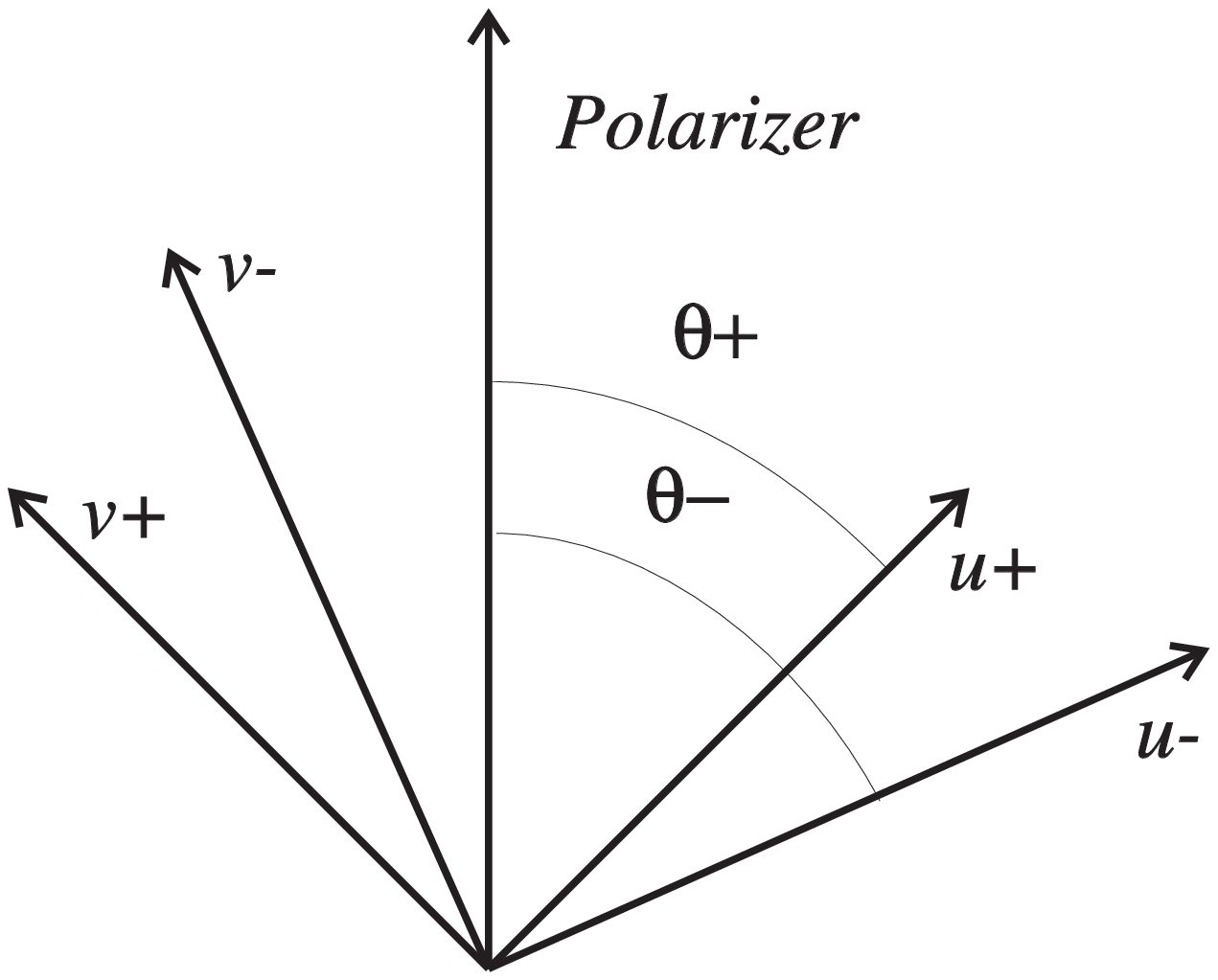}
\caption{\footnotesize  Definition of the angles $\theta_-$ and $\theta_+$.}
\label{angle}
\end{center}
\end{figure}

Figure \ref{angle} shows the respective orientation of the eigenaxis of the fiber before and after the polarizer (optical isolator).
$(u_-,v_-)$ are the field components just before the polarizer, $(u_+,v_+)$ just after it.
The effect of the polarizer can be written as
\bq \left(\ba{c}u_+\\v_+\ea\right)=\beta\left(\ba{c}\cos\theta_+\\\sin\theta_+\ea\right)
\left(\cos\theta_-\;\;\;\sin\theta_-\right)\left(\ba{c}u_-\\v_-\ea\right),\label{6}\eq
where $\beta$ is its transmission coefficient  ($\beta < 1$). Immediately after the polarizer,
 the field has a well-defined linear polarization. Denote by $f_n$ its amplitude at the beginning of the $n^{\rm th}$ round trip.
Equation (\ref{5}) can be written as
\bq
f_{n+1}=\beta\left(\cos\theta_-\;\;\;\sin\theta_-\right)
\left(\ba{c}u_ {-,n}\\v_{-,n}\ea\right),\label{7}\eq
and
\bq \left(\ba{c}u_{+,n+1}\\v_{+,n+1}\ea\right)
=\left(\ba{c}\cos\theta_+\\\sin\theta_+\ea\right)f_{n+1},\label{8}\eq
where $(u_n, v_n)$ are the field components during the $n^{\rm th}$ round.
The field components $u_{+n}=u_n(0)$ and $v_{+n}=v_n(0)$, at the entrance of the fiber immediately after the polarizer,
are transformed respectively into $u_n(L)=u_{-n}$ and $v_n(L)=v_{-n}$, at
the end of the fiber immediately before the polarizer
($L$ is the fiber length).

$u_n(0)$ and $v_n(0)$ are proportional to $f_n$. Using the propagation formulas
(\ref{4}) and (\ref{5}), $u_n(L)$ and $v_n(L)$ can be computed.
Then $f_{n+1}$ is computed as a function of $f_n$ using (\ref{7}). This yields
\bq
f_{n+1}=\beta e^{gL}\left\{Q
f_n
+\e\left[\left(\rho-\frac{i\beta_2}2\right)L
Q\partial_t^2f_n
+i P f_n\left|f_n\right|^2\right]\right\}
+O\left(\e^2\right),\label{9}\eq
with
\bq Q=\cos(\theta_+-\theta_-)\cos KL-i\cos(\theta_++\theta_-)\sin KL,\label{10}\eq
and
\bq
\ba{r}
P=
\gamma\displaystyle\biggl(\frac{e^{2gL}-1}{2g}\biggl[
\cos(\theta_+-\theta_-)\cos KL-i\cos(\theta_++\theta_-)\sin KL\hspace{1.1cm}\vspace{1.5mm}\\
\displaystyle+\frac{A-1}2\sin2\theta_+
\left[\sin(\theta_++\theta_-)\cos KL-i\sin(\theta_+-\theta_-)\sin KL\right]\biggr]
\hspace{.5cm}\vspace{1.5mm}\\
\displaystyle + \frac B2\sin2\theta_+\biggl[\sin\theta_+\cos\theta_-e^{-iKL}
\frac{e^{(2g+4iK)L}-1}{2g+4iK}\hspace{.5cm}\vspace{1.5mm}\\
\displaystyle
+\cos\theta_+\sin\theta_-e^{iKL}
\frac{e^{(2g-4iK)L}-1}{2g-4iK}\biggr]\biggr).\ea\label{11}\eq

\subsubsection{Gain threshold and continuous limit for $f_n$}

The dominant behavior of $\left(f_n\right)$ is given by the zero
order term in (\ref{9}):
\bq f_{n+1}=\beta e^{gL}Q
f_n
+O\left(\e\right).\label{12}\eq
A steady state over a large number of round trips could be reached if this first order evolution
is the multiplication by a phase factor, but only in this case.
Therefore the modulus of $\beta e^{gL}Q$ must be 1. This gives the threshold gain value:
\bq\ba{rl}g_0=&\displaystyle\frac{-1}{2L}\ln\left(\beta^2\left|Q\right|^2\right)\vspace{1.5mm}\\
=&\displaystyle
\frac{-1}{2L}\ln\left(\beta^2\left[\cos^2(\theta_+-\theta_-)
-\sin2\theta_+\sin2\theta_-\sin^2KL\right]\right)
.\ea\label{13}\eq
Denote by $e^{i\psi}$ the quantity $\beta e^{g_0L}Q$.
The exact condition is  $\left|\beta e^{gL}Q\right|=1+O\left(\e\right)$,
the gain $g$  therefore writes $g=g_0+\e g_1$, in which $g_1$ is still free.
Expanding $e^{\e g_1L}$ in a power series of $\e$, equation (\ref{9}) becomes
\bq
f_{n+1}=e^{i \psi}\left(1+\e g_1L\right)f_n
+\e\left(\rho-\frac{i\beta_2}2\right)Le^{i \psi}\partial_t^2f_n
+i \e \frac{Pe^{i\psi}}Q
 f_n\left|f_n\right|^2+O\left(\e^2\right).\label{14}\eq

Consider now some function $f$ of a continuous variable $z$, obeying an equation of the form
\bq
i\partial_zf=\left({\cal A}+i\e{\cal B}\right)f+\e{\cal C}\partial_t^2f
+\e{\cal D}f\left|f\right|^2,\label{15}\eq
where $\cal A$ and $\cal B$ are real, and $\cal C$ and $\cal D$ complex coefficients.
We look for solutions under the form $f=f_0+\e f_1+\ldots$. The solution at first order is
\bq f_0=\tilde f_0 e^{-i{\cal A}z}.\label{16}\eq

We write
\bq f_1=\tilde f_1(z,t) e^{-i{\cal A}z},\label{16b}\eq
and obtain:
\bq f(L)=\left[f(0)+\e\left({\cal B}f(0)-i{\cal C}\partial_t^2f(0)-
i{\cal D}f(0)\left|f(0)\right|^2\right)L\right]e^{-i{\cal A}L}+O\left(\e^2\right).
\label{17}\eq
Equations (\ref{14}) and (\ref{17}) are  identified according to
$f(0)\equiv f_{n}$ and $f(L)\equiv f_{n+1}$,to yield
\bq e^{-i{\cal A}L}=e^{i\psi},\label{18}\eq
\bq {\cal B}=g_1,\label{19}\eq
\bq {\cal C}=\frac{\beta_2}2+i\rho,\label{20}\eq
\bq {\cal D}=\frac{-P}{QL}.\label{21}\eq

Then the continuous equation (\ref{15}) yields some interpolation of
the discrete sequence $\left(f_n\right)$.
£££
The continuous approximation is relevant when the number of round trips is very large;
 further, mode-locked pulses actually correspond to this situation.
 The small correction of order $\e$  in equations (\ref{15}) or (\ref{17})
  gives account for the  variations of $f$ on `propagation distances', which are here number of round trips,
 very large, of order $1/\e$.
 This can be shown in a  rigorous way using the multiscale formalism, commonly used for  the derivation of
 the model equations in the soliton theory \cite{tan67a}.
 We introduce a slow variable $\zeta=\e z$, in such a  way that
\bq\partial_z=\partial_{\hat z}+\e \partial_{\zeta}.\label{22}\eq
The  values of $\zeta$ about 1 correspond to number of round trips about $1/\e$.
$f$ is expanded as above as $f_0+\e f_1+\ldots$, and the
first order obviously yields (\ref{16}), we make the transform (\ref{16b})
and $\tilde f_1$ must satisfy the following equation:
\bq i\partial_{\zeta}\tilde f_0+i\partial_{\hat z}\tilde f_1=
i{\cal B}\tilde f_0+{\cal C}\partial_t^2\tilde f_0+
{\cal D}\tilde f_0\left|\tilde f_0\right|^2.\label{23}\eq
Equation (\ref{23}) can be written as
\bq \partial_{\hat z}\tilde f_1={\cal F}(\tilde f_0).\eq
The long distance evolution of $\tilde f_0$ is obtained from the requirement
that $\tilde f_1$ does not grow linearly with $z$.
This yields $\partial_{\hat z}\tilde f_1=0$ and
\bq i\partial_{\zeta}\tilde f_0=
i{\cal B}\tilde f_0+{\cal C}\partial_t^2\tilde f_0+
{\cal D}\tilde f_0\left|\tilde f_0\right|^2.\label{24}\eq
Or, using real coefficients only:
\bq i\partial_{\zeta}\tilde f_0=
i g_1\tilde f_0+\left(\frac{\beta_2}2+i\rho\right)\partial_t^2\tilde f_0+
\left({\cal D}+i{\cal D}_i\right)\tilde f_0\left|\tilde f_0\right|^2.\label{24bis}\eq
Equation (\ref{24}) or (\ref{24bis}) is the cubic complex Ginzburg-Landau (CGL) equation. Note that equation
(\ref{24}) is formally identical to the master equation proposed by Haus {\it et al} \cite{7}. However,
 the coefficients of (\ref{24}) explicitly depend on the orientation of the eigenaxis of the fiber at both sides of the polarizer.
An essential feature is the arising of a nonlinear gain or absorption ${\cal D}_i$, which results from the combined effects of
 the nonlinear rotation of the
polarization, the losses due to the polarizer, and the linear gain. The value and the sign of  ${\cal D}_i$ depend on the
angles $\theta_+$ and $\theta_-$ between the eigenaxis of the fiber and the polarizer.

\subsection{Different regimes for  the  CGL equation}
\subsubsection{The stationary solution and its modulational instability}
The CGL equation (\ref{24bis}) admits  the following nonzero stationary solutions, {\it ie} with a constant modulus:
\bq f_1=A e^{i\left(\kappa\zeta-\Omega t\right)}\label{n1}\eq
with
\bq \Omega^2=\frac1\rho\left({\cal D}_i\vert A\vert^2+g_1\right)\label{n2}\eq
and
\bq \kappa=\frac{\beta_2}{2\rho}\left({\cal D}_i\vert A\vert^2+g_1\right)-{\cal D}_r\vert A\vert^2.\label{n3}\eq
In the particular case $\Omega=0$ it is a constant solution (independent of $t$),
with a fixed amplitude
\bq A=\sqrt{\frac{-g_1}{{\cal D}_i}},\label{n4}\eq
and \bq \kappa=\frac{{\cal D}_r}{{\cal D}_i}g_1.\label{n5}\eq
It exists only if ${\cal D}_ig_1<0$, {\it ie} if the excess of linear gain $g_1$
 and the effective nonlinear gain ${\cal D}_i$ have opposite signs.
We perform a linear stability analysis for the constant  solution $f_c=Ae^{i\kappa\zeta}$.
We seek for solutions of the form $f_1=f_c\left(1+u\right)$, $u$ being very small. It is seen that it must satisfy the
following equation:
\bq -\kappa u+i\partial_\zeta u=ig_1 u+\left(\frac{\beta_2}2+i\rho\right)\partial_t^2u+{\cal D}A^2\left(2u+u^\ast\right).\label{n6}\eq
 Due to the term $u^*$, equation (\ref{n6}) is not linear.
The eventual instability is of the type discovered  first by Benjamin and Feir in the frame of water waves \cite{ben67}.
It has been shown  that this kind of instability  involves
the nonlinear interaction of  two modulation terms
with conjugated phases with the square $A^2$ of the constant solution \cite{stu78}.
Thus we write $u$ as:
\bq u=u_1e^{\left(\lambda \zeta-i\omega t\right)}+u_2e^{\left(\lambda^\ast \zeta+i\omega t\right)}\label{n7}\eq
It is found that $\lambda$ satisfies an equation of the form
\bq \lambda^2+2b\lambda+c=0.\label{n8}\eq
Using the expression (\ref{n4}) of the fixed amplitude $A$, we get the following values of the real constants
$b$ and $c$:
\bq b=g_1+\rho\omega^2,\label{n9}\eq
\bq c=\left(g_1\frac{{\cal D}_r}{{\cal D}_i}+\frac{\beta_2}2\omega^2\right)^2+b^2-
g_1^2\left(1+\frac{{\cal D}_r^2}{{\cal D}_i^2}\right).\label{n10}\eq
A short analysis of equation (\ref{n8}) show that a necessary condition for $u$ to remain bounded is that
$b$ is positive. This proves that modulational instability occurs as soon as the excess of linear gain $g_1$ is negative
and the nonlinear gain ${\cal D}_i$ is positive. Recall indeed that the existence of the constant nonzero solution
requires that these two quantities have opposite signs.
When $g_1>0$ and ${\cal D}_i<0$, we show that the modulational instability never occurs.
Numerical computation shows indeed that the quantity $\left({\cal D}_r\beta_2/2+{\cal D}_i\rho\right)$ is always negative,
which ensures, using elementary analysis,  that $c$ is always positive, and the result follows.

\begin{figure}[hbt!]
\begin{center}
\includegraphics[width=9cm]{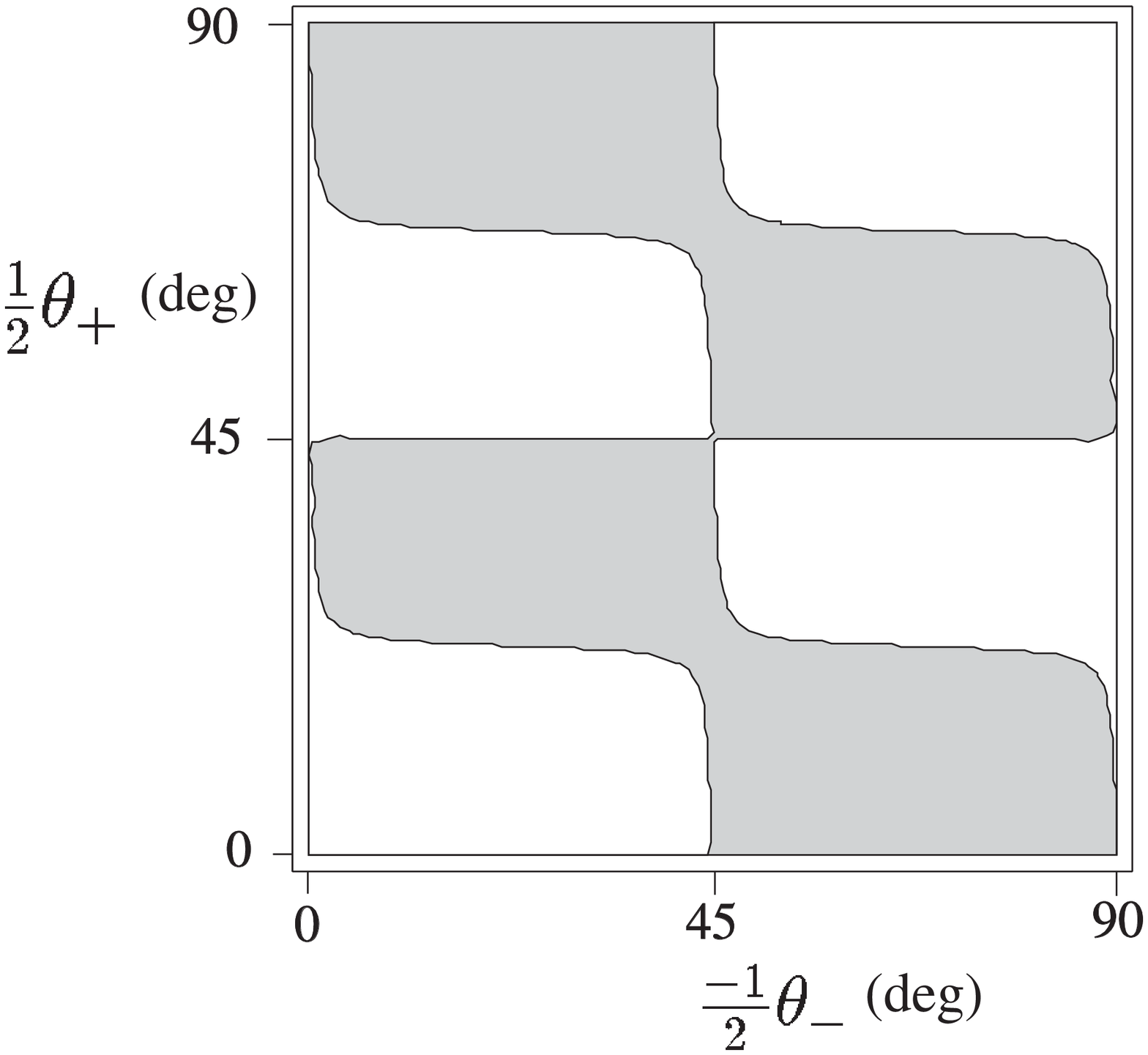}
\caption{\footnotesize  Sign of the effective nonlinear gain ${\cal D}_i$ as a function of the orientation
of the  polarizer. White: ${\cal D}_i<0$, and the constant solution of the CGL equation is stable. Gray:
 ${\cal D}_i>0$, and the constant solution of the CGL equation is unstable.}
\label{ddi}
\end{center}
\end{figure}

If we admit that the excess of linear gain $g_1$ will self-adjust to a value for which a stable solution exists,
the above discussion shows that the  stability of the constant  solution depends on the sign of the
effective nonlinear gain ${\cal D}_i$.
It is stable when ${\cal D}_i<0$ only.
 We expect to observe continuous laser emission when the constant solution is stable,
and only in this case. Thus the domains of continuous emission should coincide with the regions where ${\cal D}_i$ is
negative. The sign of ${\cal D}_i$ as function of the angles  $(\frac{-1}2\theta_-,\frac12\theta_+)$ is  drawn on figure \ref{ddi}.
 It  is easily checked that the relation between the angles $\theta_1$, $\theta_2$ used in the experiments and the
 angles $\theta_+$, $\theta_-$ defined by figure \ref{angle} is such that $\theta_1=\frac{-1}2\theta_-+\theta_{10}$ and
$\theta_2=\frac{1}2\theta_++\theta_{20}$, with some fixed value of $\theta_{10}$  and $\theta_{20}$.
The white domain on figure \ref{ddi} corresponds to the negative values of the nonlinear gain ${\cal D}_i$. It is thus the domain where the constant solution
of the CGL equation is stable. Continuous laser emission should occur in this domain.
\subsubsection{Localized solutions}
\begin{figure}[hbt!]
\begin{center}
\includegraphics[width=9cm]{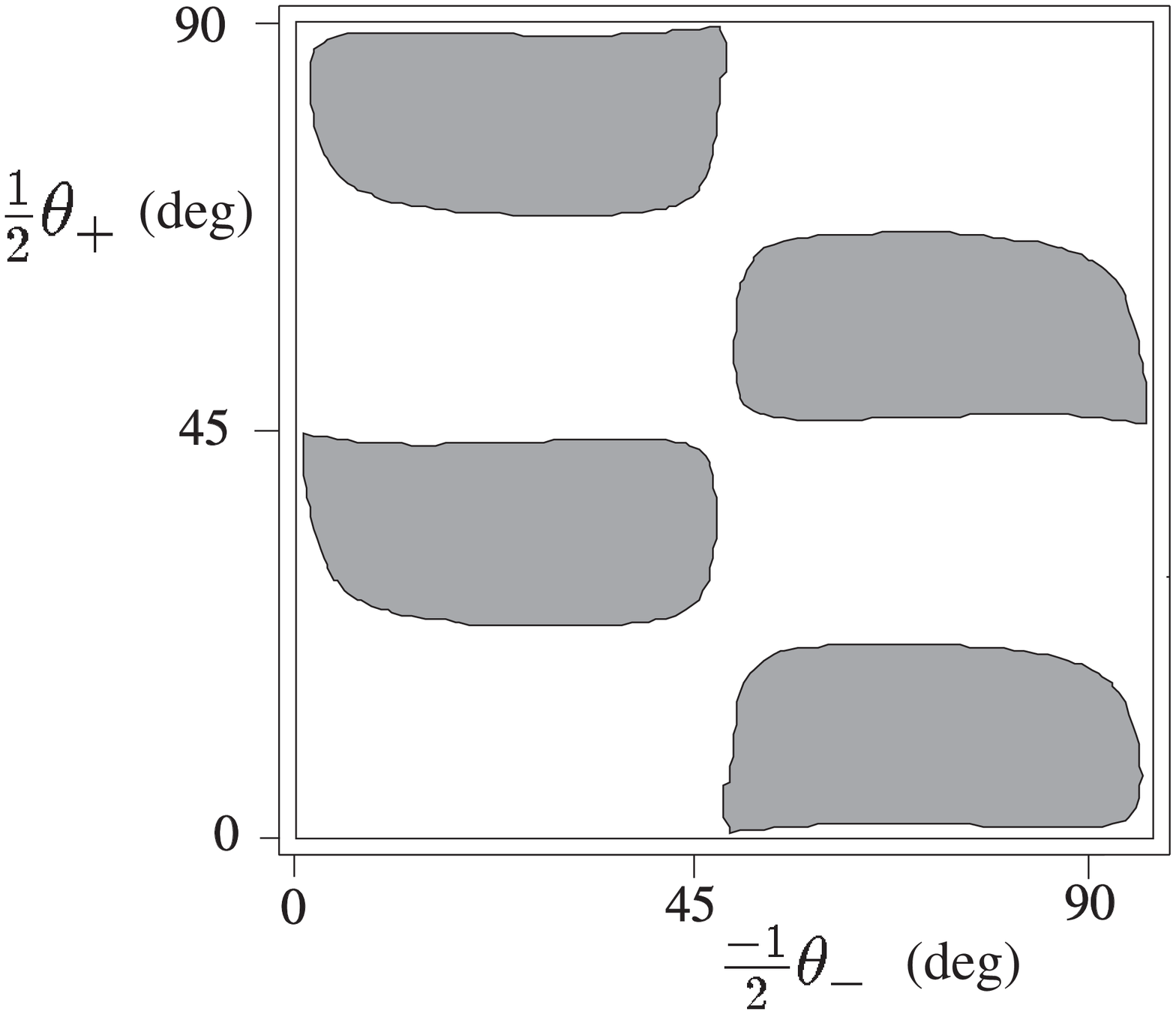}
\caption{\footnotesize  Sign of the excess of nonlinear gain value $g_1$ for which the exact localized solution exists,
as a function of the orientation
of the  polarizer. White: ${\cal T}$ and $g_1$ positive. Gray:
 ${\cal T}$ and $g_1$ negative.}
\label{pltruc2}
\end{center}
\end{figure}

\begin{figure}[hbt!]
\begin{center}
\includegraphics[width=9cm]{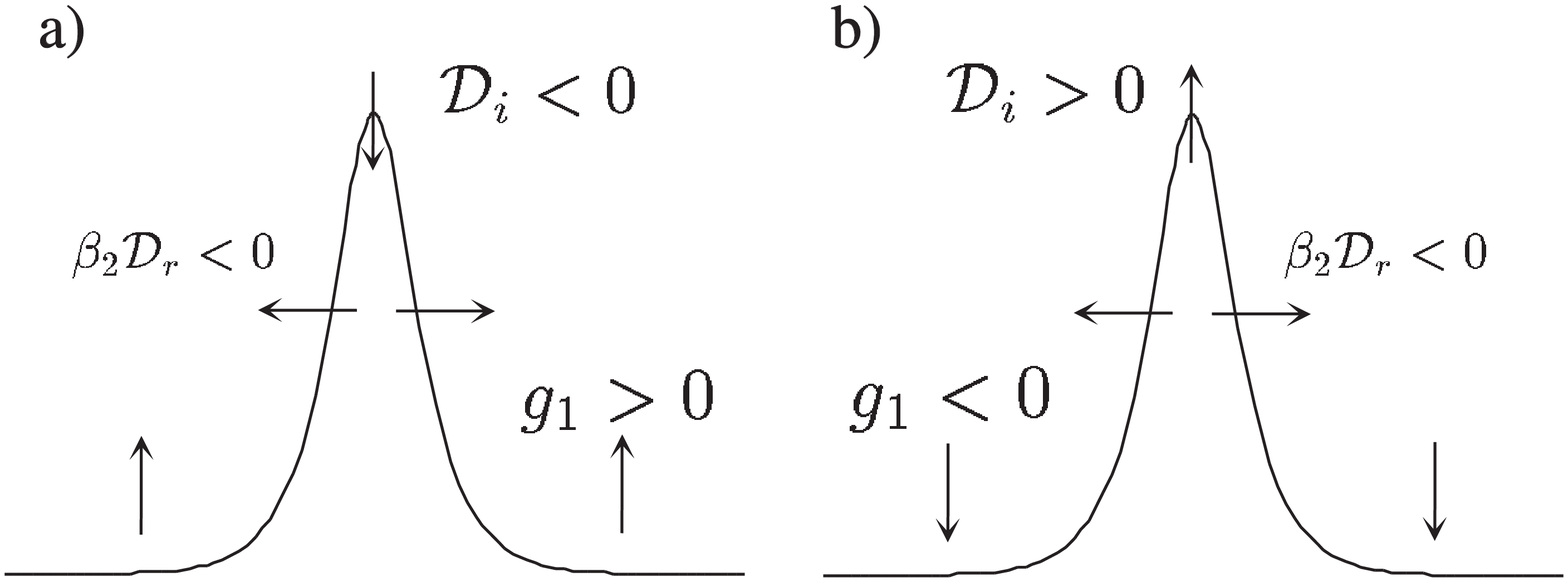}
\caption{\footnotesize  Schematic representation of the effect of the nonlinear gain ${\cal D}_i$, of the excess of
linear gain $g_1$,  and of both the dispersion $\beta_2$ and the effective self-phase modulation ${\cal D}_r$ on a localized pulse.}
\label{pulse}
\end{center}
\end{figure}

A localized analytical solution of the CGL equation (\ref{24bis}) can also be written \cite{sot97a}. It has the following expression:
\bq f_1=a(t) e^{i\left(d \ln a(t)-\omega \zeta\right)}\label{n11},\eq
with \bq a(t)=BC\mbox{sech }\left(Bt\right)\label{n12}\eq
and \bq B=\sqrt{\frac{g_1}{\rho d^2-\rho-\beta_2 d}},\label{n13}\eq
\bq C=\sqrt{\frac{3d\left(4\rho^2+\beta_2^2\right)}{2\left(\beta_2 {\cal D}_i-2\rho {\cal D}_r\right)}},\label{n14}\eq
\bq d=\frac{3\left(\beta_2{\cal D}_r+2\rho {\cal D}_i\right)+\sqrt{9\left(\beta_2{\cal D}_r+2\rho {\cal D}_i\right)^2+8
\left(\beta_2{\cal D}_i-2\rho {\cal D}_r\right)^2}}{2\left(\beta_2{\cal D}_i-2\rho {\cal D}_r\right)},\label{n15}\eq
\bq\omega=\frac{-g_1\left(4\rho d+\beta_2 d^2-\beta_2\right)}
{2\left(\rho d^2-\rho-\beta_2 d\right)}.\label{n16}\eq
The inverse $B$ of the  pulse length is real only if the quantity ${\cal T}=\left(\rho d^2-\rho-\beta_2 d\right)$ and the
 excess of linear gain $g_1$ have the same sign.
The regions where ${\cal T}$ is either positive or negative are specified on figure \ref{pltruc2}.
The background with zero amplitude is stable when the excess of linear gain $g_1$ is negative, and unstable in the opposite case.
When $g_1>0$, the exact localized solution (\ref{n11}) is unstable due to the instability of the background \cite{sot97a}.
Qualitatively, localized pulse formation can be expected
when the excess of linear gain $g_1$ is negative  and the nonlinear gain ${\cal D}_i$ is positive, as is suggested on figure~\ref{pulse}.
The effective self-phase modulation ${\cal D}_r$  is always negative,  and the dispersion $\beta_2$ is positive.
Therefore their conjugated effect
 leads to the increase of the pulse width.
If $g_1$ is positive and ${\cal D}_i$ negative (figure \ref{pulse}a), the nonconservative effects decrease the amplitude at the top of the pulse, and
increase it at the bottom: no stable localized pulse can be formed. If on the contrary $g_1$ is negative and ${\cal D}_i>0$
(figure \ref{pulse}b),
 energy appears at the
top of the pulse and disappears at its bottom, yielding some pulse narrowing, which could be expected to balance the
broadening caused by the nonlinear index variations.
In fact the analytical localized solution (\ref{n11}) is never stable when $\beta_2 {\cal D}_r<0$, but can be
stabilized when higher order  nonlinear terms are taken into account, as shows the study of the so-called quintic CGL equation
 \cite{sot97a}. The stabilizing term should be a quintic nonlinear absorption.
Observe that here the effective nonlinear gain ${\cal D}_i$
 follows from the linear gain and the nonlinear evolution of the polarization in the fiber, without
 considering cubic nonlinear gain in   the initial model
(\ref{1}-\ref{2}). In an analogous way, an effective
 quintic nonlinear gain  term in the master equation (\ref{24}) does not require that a quintic nonlinear gain  term
is present  in   the initial model, either a cubic nonlinear gain or a quintic nonlinear index should be enough.
(\ref{1}-\ref{2}).
 The precise determination of this higher order term
is  left for further study. On these grounds, the  exact solution (\ref{n11})  corresponds to a potentially stable pulse
when $g_1 <0$ and ${\cal D}_i>0$, and to a completely unstable pulse when  $g_1 >0$ and ${\cal D}_i<0$.

\subsubsection{The different regimes}
\begin{figure}[hbt!]
\begin{center}
\includegraphics[width=9cm]{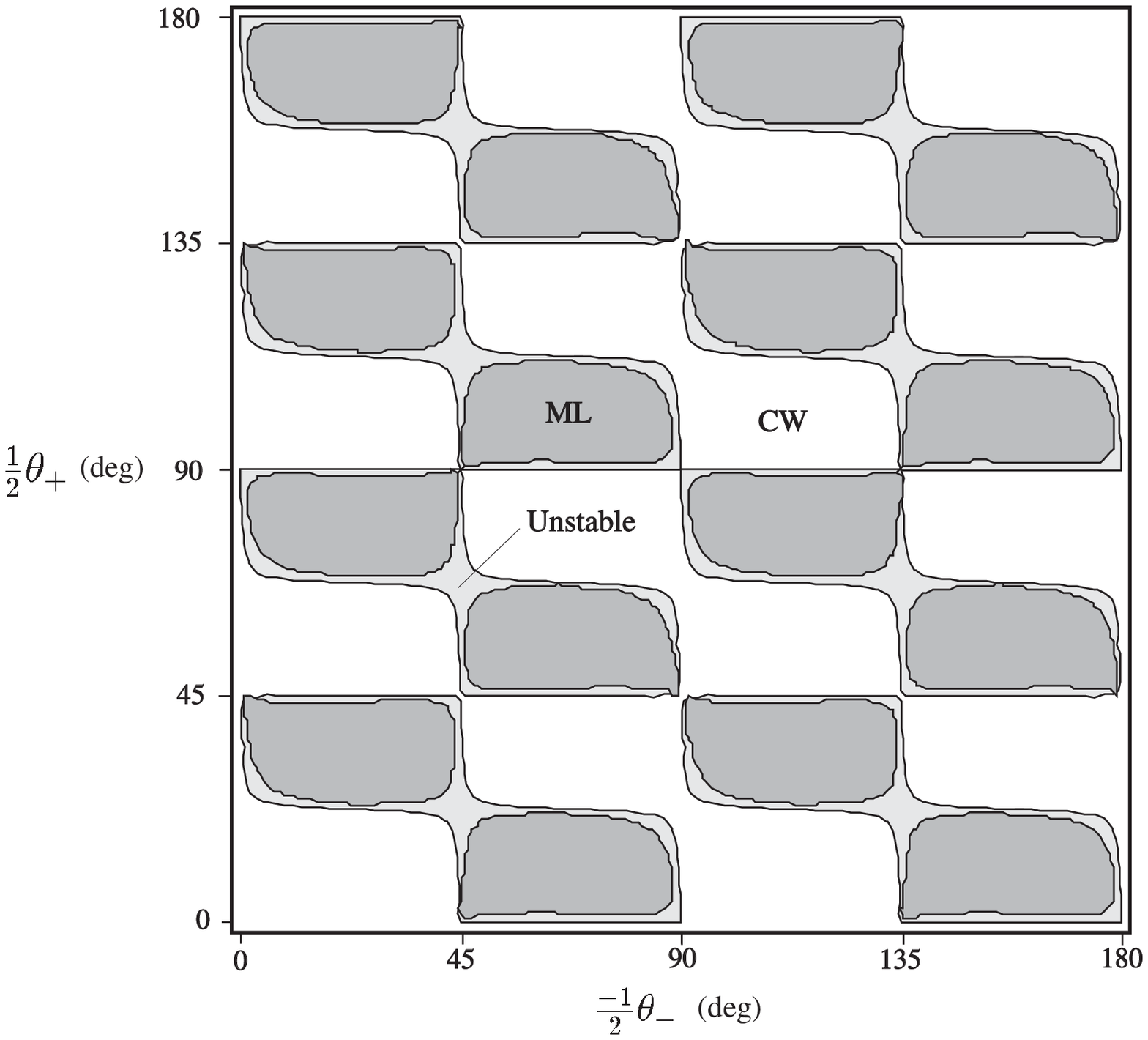}
\caption{\footnotesize  Operating regime of the laser as a function of the angles $\theta_+$ and $\theta_-$ between the polarizer and the
eigenaxis of the fiber, according to the theory. ML stands for mode-locked regime and CW for continuous wave regime.}
\label{concl}
\end{center}
\end{figure}

According to the conclusions of the previous section, and assuming that
 the excess of linear gain $g_1$ will self-adjust
to the value for which a stable solution exists, mode-locked laser emission can be expected
in the region where  ${\cal D}_i>0$ and ${\cal T}<0$. This is the domain in dark gray marked ML on figure \ref{concl}.
 In the region where  ${\cal D}_i$ is negative, the constant solution of the CGL equation is stable, and
 continuous laser emission is expected to occur. It is the white domain  marked cw on  the figure \ref{concl}.
   In the region where ${\cal T}$ is positive and ${\cal D}_i$ negative,
no stable localized  neither constant
solution exist. The laser behavior is expected to be unstable in this region, in light gray on the figure \ref{concl}.
The theoretical results summarized on the figure \ref{concl} can be compared to the experimental results of figure \ref{figs2}.
A discrepancy   between the two figures appears at first glance: the periodicity of the
behavior regarding the variable $\theta_2$ or $\theta_+/2$ seems to be  45 degrees according to the theory,
 while it is  90 degrees according to  the experiment.  Indeed,
 the two elongated domains of mode-locked behavior are not equivalent according to  experimental observation,
 while they seem to be identical on the theoretical figure \ref{concl}.
\begin{figure}[hbt!]
\begin{center}
\includegraphics[width=9cm]{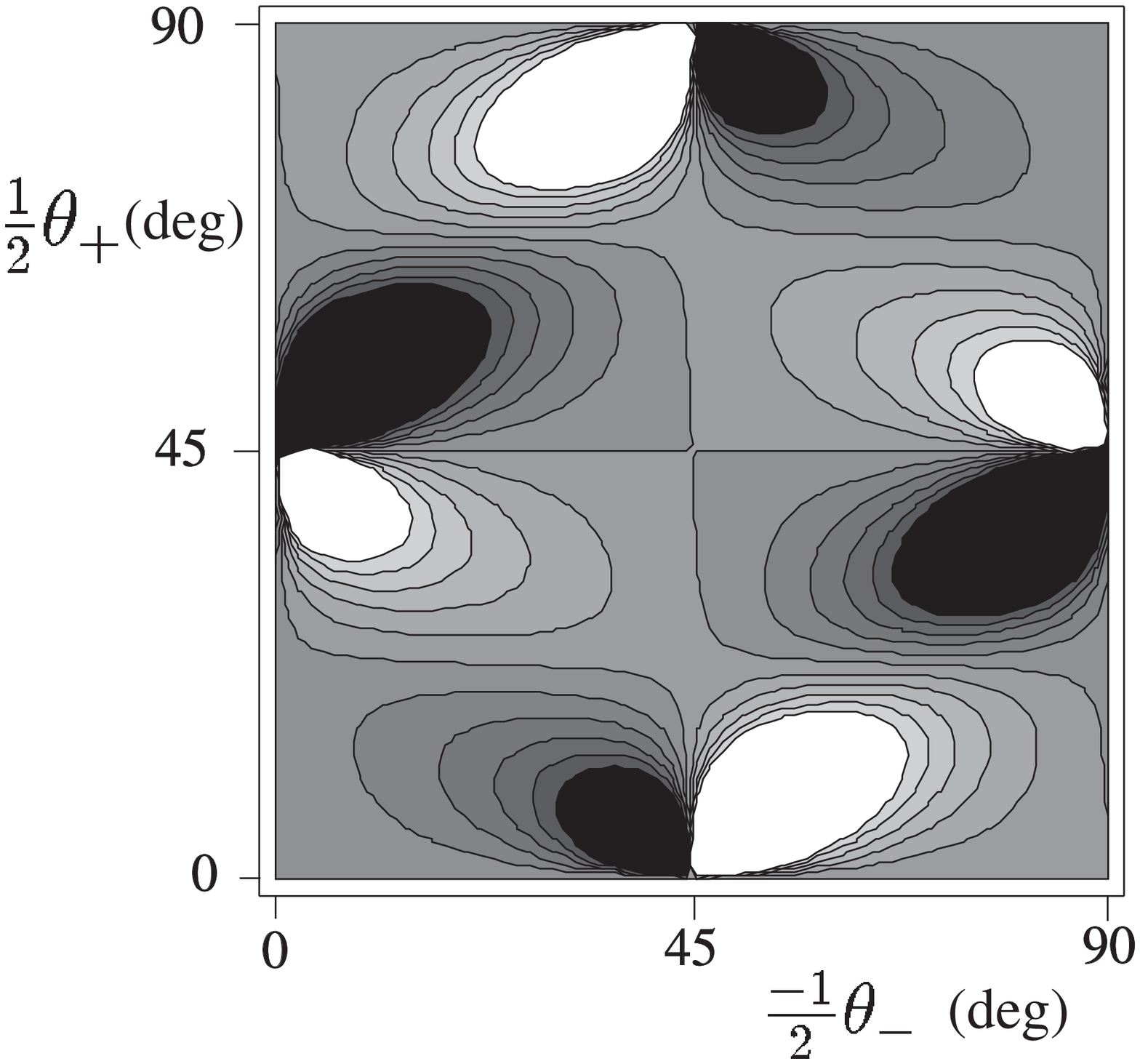}
\caption{\footnotesize  Contour plot of the effective nonlinear gain ${\cal D}_i$
 as a function of the angles $\theta_+$ and $\theta_-$ between the polarizer and the
eigenaxis of the fiber. }
\label{codi}
\end{center}
\end{figure}
A contour plot of the effective nonlinear gain ${\cal D}_i$  as a function of the angles is drawn on figure \ref{codi}.
It is seen that the values of  ${\cal D}_i$ are not the same in the two theoretical domains of  mode-locked emission.
Thus they are in fact not equivalent, and the periodicity of the theoretical result is   90 degrees, as that of the
experimental one. Further it has been shown \cite{sot97a} that an excessive value of the nonlinear gain might
 prevent  pulse stabilization.
A more accurate analysis would thus very likely show that
the  regions where ${\cal D}_i$ takes its
 highest values are out of the domain of  stability of the localized pulse.
 This confirms that the two horizontal domains of mode-locked regime drawn on figure \ref{concl} are not equivalent.
 Further, it  could explain a part of the discrepancy between theoretical and experimental  results.

Before we conclude, let us consider the influence of the group velocity dispersion
on the stability of the mode-locking solutions. This is important because GVD compensation
 is required to generate subpicosecond pulses. Experimentally, we have obtain
 666 fs pulses with a grating pair inserted in the cavity \cite{6,11bis}.
 Under these conditions, the total GVD is about $\beta_2 = 0.0005\,\rm  ps^2/m$.
  Hence, stable ultrashort pulse generation is possible for such low values of the GVD.
  We have tested our master equation for decreasing values of $\beta_2$.
   We have found that a lowering of $\beta_2$ results in a reduction of the
    stable mode-locking regions. In addition, mode-locking completely disappears for
    $\beta_2\lesssim 0.0015 \,\rm ps^2/m$.
     These results are in agreement with the predictions of the Haus's model \cite{7}
      but not with our experimental results.
       The validity of the model presented here is therefore limited to GVD values not
       too close to zero. Higher order terms of the GVD could be required to improve the
        model near the zero-dispersion point. This problem deserves further experimental
         and theoretical work.

\section{Conclusion}
We have given a rigorous derivation of the master equation, a CGL equation, that allows to discuss
theoretically  the different operating regimes of the passively mode-locked Yb-doped double-clad fiber laser.
 This equation involves a single amplitude, but gives an account
of the evolution of the polarization inside the fiber through its coefficients. An effective  nonlinear
gain arises from the conjugated effects of the linear gain and the self-phase modulation. Its
sign depends on the orientation of the polarizer.
The existence and stability of the  constant nonzero solution of the CGL equation has been discussed. It gives
an account of the continuous emission regime of the laser. An analytical localized  solution has also been given.
Although it is properly speaking unstable, it allows to discuss the formation of localized pulses, {\it ie} the
mode-locked regime of the laser. An unstable regime has  also been identified. These theoretical results are in good accordance
with the experiments. The study of higher order corrections to this model should  improve these results.

\end{document}